\documentclass[12pt]{JHEP3}
\usepackage{mathrsfs}
\usepackage{amsmath,amssymb}
\usepackage{epsfig}
\usepackage{relsize}
\usepackage{graphicx}

\def\hK{\hat{K}}

\def\tH{\widetilde{{\cal E}}}

\def\sl(2){\alg{sl}(2)}

\def \F {{\cal F}}
\def \cs {{\alg s}}

\def\be{\begin{equation}}
\def\ee{\end{equation}}

\newcommand{\bea}{\begin{eqnarray}}
\newcommand{\eea}{\end{eqnarray}}

\def\br {\bar{\rho}}
\def\a {\alpha}

\def\g {\gamma}
\def\om {\omega}

\def\la{\label}
\def\e{\epsilon}
\def\ov{\over}

\def\tp{{\widetilde p}}

\newcommand{\alg}[1]{\mathfrak{#1}}
\newcommand{\su}{\alg{su}}

\newcommand{\AdS}{{\rm  AdS}_5\times {\rm S}^5}
\newcommand{\ads}{{\rm  AdS}_5\times {\rm S}^5}

\newcommand{\atopfrac}[2]{\genfrac{}{}{0pt}{}{#1}{#2}}

\newcommand{\bem}{\left (\begin{matrix}}
\newcommand{\eem}{\end{matrix} \right )}

\author{Gleb Arutyunov$^a$\footnote{Email: G.E.Arutyunov@uu.nl, frolovs@maths.tcd.ie} {}\footnote{Correspondent fellow at Steklov
Mathematical Institute, Moscow.}\, and\,  Sergey Frolov$^{b\, \dagger}$
 \\ $^{a}$ {\it Institute for Theoretical
Physics and Spinoza Institute,\\ ~~Utrecht University, 3508 TD
Utrecht, The Netherlands} \\ $^b$ {\it Hamilton Mathematics Institute and School of Mathematics, \\
~~Trinity College, Dublin 2,
Ireland} }

\abstract{ We use the string hypothesis for the mirror theory to
derive the Thermodynamic Bethe Ansatz equations for   the $\AdS$
mirror model. We further demonstrate how these equations can be
used to construct the associated Y-system recently discussed in
the literature, putting particular emphasis on the assumptions and
the range of validity of the corresponding construction.}

\title{Thermodynamic Bethe Ansatz \\ for  the  $\AdS$
Mirror Model}

\preprint{
          \smaller{\smaller{\smaller{ITP-UU-09-09}}}\\[-.5ex]
          \smaller{\smaller{\smaller{SPIN-09-09}}}\\[-.5ex]
          \smaller{\smaller{\smaller{TCDMATH 09-09}}}\\[-.5ex]
          \smaller{\smaller{\smaller{HMI-09-05}}}}

\begin{document}

\renewcommand{\thefootnote}{\arabic{footnote}}
\setcounter{footnote}{0}


\section{Introduction}

Recently, there has been further important progress towards
understanding the finite-size spectrum of the $\AdS$ superstring.
The conjectured quantum integrability of the string model plays a
primal role in those developments.

\smallskip

In the uniform light-cone gauge the string sigma model is
described by a two-dimensional massive integrable quantum field
theory defined on a cylinder of circumference equal to the
light-cone momentum $P_+$; the latter can be viewed as the length
$L\equiv P_+$ of the string, see the review \cite{AFrev}. In the
limit of infinite $L$ the spectrum of the model is known to
contain a set of elementary particles which transform in the
bi-fundamental representation of the centrally extended
superalgebra $\su(2|2)$. In addition, there are $Q$-particles,
which can be thought of as bound states of $Q$ elementary
particles \cite{D1}. The $Q$-particles reside into the tensor
product of two $4Q$-dim atypical totally symmetric multiplets of
the centrally extended $\su(2|2)$ algebra. In the infinite $L$
limit the symmetries of the model are powerful enough to determine
the matrix structure of both the fundamental S-matrix
\cite{B,AFZzf} and the S-matrices of $Q$-particles \cite{AFb,
ALT}. Crossing symmetry puts further constraints \cite{Janik} on
the normalizations of all the S-matrices.

\smallskip

When $L$ is large but finite, the multi-particle states can be
approximately described by the wave function of the Bethe-type
\cite{ZZ}. Factorizability of the multi-particle scattering
together with the periodicity condition for the wave function then
implies the quantization conditions for the particle momenta which
are encoded into a set of the Bethe-Yang equations. These
equations  \cite{BS} give a proper (asymptotic) description of the
string spectrum for $L$ large, but they become less and less
accurate when the value of $L$ decreases. The effect of a finite
volume manifests itself in the appearance of exponentially small
corrections to the particle energies computed via the Bethe-Yang
equations. Derivation of the leading corrections by means of
L\"uscher's approach \cite{Lu} has been in the focus of recent
investigations \cite{AJK}-\cite{Beccaria:2009eq}.

\smallskip

To obtain the finite $L$ spectrum, one could try to generalize the
Thermodynamic Bethe Ansatz approach (TBA) originally developed for
relativistic integrable models \cite{za}. To this end, one starts
with considering the string sigma model on a torus given by the
Cartesian product of two circles with circumferences $L$ and $R$,
respectively. In the imaginary time formalism, the circumference
of any of these two circles can be regarded as the inverse
temperature in a statistical field theory with the Hilbert space
of states defined on the complementary circle. Interchanging the
role of the two-dimensional space and time, and further taking the
limit $R\to \infty$, allows one to relate the ground state energy
in the original string model with the free energy (or, depending
on the boundary conditions for fermions, with Witten's index) in
the so-called {\it mirror} model; the latter is obtained from the
original theory by a double Wick rotation. It has been shown that
the TBA approach is also capable of accounting for the excited
states \cite{BLZe,DT}, see \cite{Fendley:1997ys}-\cite{Ahn:2003st}
for further results and different approaches, and also \cite{GKV}
for the recent important development.

\smallskip

If a theory is relativistic, then it coincides with its mirror.
The light-cone gauge string sigma model lacks two-dimensional
Lorentz invariance and, for this reason, the corresponding mirror
model is a new theory that requires a separate exploration. In our
previous work \cite{AFtba} the Bethe-Yang equations for elementary
particle of the mirror theory have been derived. Also, the bound
states of elementary particles have been classified and shown to
comprise into the tensor product of two $4Q$-dimensional atypical
totally anti-symmetric multiplets of the centrally extended
$\su(2|2)$ algebra. The Bethe-Yang equations for the corresponding
$Q$-particles are easily obtained by fusing the elementary
equations\footnote{These equations also follow from the
requirement of Yangian symmetry \cite{deLeeuw:2008ye}.}. More
recently,
 we noticed
\cite{AFsh} that the mirror Bethe-Yang equations involving
auxiliary roots can be interpreted as the Lieb-Wu equations
\cite{Lieb} for an inhomogeneous Hubbard model \cite{Martins,Korepin}.
This inhomogeneous Hubbard model becomes homogeneous in the limit
of infinite real momenta of the mirror $Q$-particles\footnote{The
relationship between the equations \cite{BS} and the Lieb-Wu
equations for the inhomogeneous Hubbard model was observed earlier
\cite{B2}. However, for the original string model there is no
value of the physical, {\it i.e.} real particle momenta for which
the Lieb-Wu equations become homogeneous.}. Further, we have
argued that the solutions of the Bethe-Yang equations contributing
in the thermodynamic limit arrange themselves into Bethe string
configurations similar to the ones appearing in the Hubbard model
and we have obtained the corresponding Takahashi-like equations
for the real centers of the string complexes. This constitutes the
string hypothesis for the $\AdS$ mirror theory \cite{AFsh} and
offers an unequivocal way to obtain the corresponding TBA
equations.

\smallskip

The aim of the present paper is to derive the TBA equations for
the $\AdS$ mirror theory. This will be  done by following the
well-established routine, {\it i.e.} by passing to the
thermodynamic limit description of the Takahashi equations in
terms of the particle-hole densities with subsequent minimization
of the free energy. By exploiting the properties of the emerging
TBA kernels, we will be able to slightly simplify the initial
system of the TBA equations for the particle pseudo-energies. Then
we will also attempt to derive the associated Y-system
\cite{Zamolodchikov:1991et}-\cite{Bazhanov:1994ft}. Opposite to
the infinite set of the coupled TBA equations, the Y-system is a
local set of equations but it is obtained at the price of ``wiping
off" a lot of information from the original TBA system. In some
cases this information can be restored by specifying the analytic
properties of the Y-system, as has been nicely demonstrated in the
recent work \cite{GKV}. We would like to stress, however, that in
our present case the success of deriving the Y-system crucially
depends on the analytic properties of the so-called dressing phase
\cite{AFS} which constitutes a part of the TBA kernel for the
$Q$-particles. Although in the original string theory the dressing
phase
 is believed to be known, both in the
strong coupling asymptotic expansion \cite{BHL} and for a finite
value of the coupling \cite{BES}, its analytic properties and the
actual expression in the mirror theory are currently  terra
incognita. To be precise, the dressing phase by \cite{BES} is represented by a double
series convergent in the region $|x_{1}^{\pm}|>1$ and
$|x_2^{\pm}|>1$, where $x_{1,2}^{\pm}$ are kinematic parameters
related to the first and the second particle, respectively. This
series admits an integral representation found by
Dorey, Hofman and Maldacena \cite{DHM} which is valid in the same
region of kinematic parameters. To determine the dressing phase in the mirror region, one has to
analytically continue the corresponding integral representation beyond
$|x_{1,2}^{\pm}|>1$.

\smallskip

The Y-system for the planar AdS/CFT correspondence \cite{M} has
been already conjectured  \cite{GKV1} based on the experience with
the classical discrete Hirota dynamics in the ${\rm O}(4)$ model.
Very recently two independent derivations of the Y-system (and the
TBA equations) have been presented \cite{Bombardelli:2009ns,GKV2}.
Although we did not attempt to make a detailed comparison of our
results with those by \cite{Bombardelli:2009ns,GKV2}, a bird's eye
survey reveals definite similarities but also certain differences
between our findings. Relegating some of our comparative remarks
to Conclusions, we would like to stress that in our opinion the
question -- what is the mirror dressing phase -- is the most
important one towards an ultimate understanding of the TBA system.

\smallskip

The paper is organized as follows. Section 2 deals with the
thermodynamic limit of the Takahashi-like equations for the mirror
model. In section 3 we derive the TBA equations. In section 4 we
partially simplify the TBA equations and discuss the construction
of the associated Y-system. In the three appendices we summarize
the most essential properties of the mirror kinematics,  the
relations and properties of the TBA kernels, and the simplified
system of the TBA equations.

\section{Integral equations for densities}\la{tbaeqs}
In the thermodynamic limit we introduce densities $\rho(u)$ of particles, and densities $\br(u)$ of holes which depend on the real rapidity variable $u$. We have the following types of densities  ($\a=1,2$)
\begin{enumerate}
\item The density $\rho_Q(u)$ of the $Q$-particles, $-\infty\le u\le \infty$
, $Q=1,\ldots,\infty$
\item The density $\rho_{y^-}^{(\a)}(u)$ of the $y$-particles with Im$(y)<0$, $-2\le u\le 2$.
The corresponding $y$-coordinate is expressed in terms of $u$ as $y=x(u)$ where $x(u)$ is defined in (\ref{xu})

\item The density $\rho_{y^+}^{(\a)}(u)$ of the $y$-particles with Im$(y)>0$, $-2\le u\le 2$. The corresponding $y$-coordinate is expressed in terms of $u$ as $y={1\ov x(u)}$

\item The density $\rho_{M|vw}^{(\a)}(u)$ of the $M|vw$-strings, $-\infty\le u\le \infty$
, $M=1,\ldots,\infty$\item The density $\rho_{N|w}^{(\a)}(u)$ of the $N|w$-strings, $-\infty\le u\le \infty$, $N=1,\ldots,\infty$\,,
\end{enumerate}
and the corresponding densities of holes.

Introducing a generalized index $i$ which runs over all the
densities, one can represent the system of integral equations
arising in the thermodynamic limit in the following compact form
\begin{eqnarray}\la{tbacom}
\rho_i(u) + \br_i(u) = {R\ov 2\pi}{d\tp_i\ov du} + K_{ij}\star\rho_j (u)\,.~~~~
\end{eqnarray}
where the momentum $\tp_i$ does not vanish only for $Q$-particles, and is  given by (\ref{pu}).  Here the summation over $j$ is assumed, and the star product is defined by the following composition law
\bea\la{starp}
K_{ij}\star\rho_j (u) = \int {\rm d}u'\, K_{ij}(u,u')\rho_j (u')\,,
\eea
where  the integration is taken over the range of $u$ specified above.
The explicit form of the kernels $K's$ is discussed below.

The star product (\ref{starp}) should be thought of as the left
action of the kernels $K's$ on $\rho_j$. In what follows we will
also need the right action which is defined as  \bea\la{starp2}
\rho_j \star K_{ji}(u) = \int {\rm d}u'\, \rho_j
(u')K_{ji}(u',u)\,. \eea

\subsection*{Equations for $Q$-particle densities}
To derive the integral equation for $Q$-particle densities, we
rewrite the Bethe-Yang equation (3.1) in \cite{AFsh} for
$Q$-particles  in terms of the function $x(u)$  given by
(\ref{xu}) \bea\nonumber 1&=&e^{i\tp_{k} R} \prod_
{\textstyle\atopfrac{l=1}{l\neq k}}
^{K^{\mathrm{I}}}S_{\sl(2)}^{Q_kQ_l}(u_{k},u_{l})
\prod_{\a=1}^{2}\prod_{l=1}^{N_{y^-}^{(\a)}}\frac{{x(u_k-i{Q_k \ov
g })-x(u_{l}^{(\a)})}}{x(u_k+i{Q_k \ov g })-x(u_{l}^{(\a)})}
\sqrt{\frac{x(u_k+i{Q_k \ov g })}{x(u_k-i{Q_k \ov g
})}}\,\\\la{BEQ}
&\times&\prod_{\a=1}^{2}\prod_{l=1}^{N_{y^+}^{(\a)}}\frac{{x(u_k-i{Q_k
\ov g })-{1\ov x(u_{l}^{(\a)})}}}{x(u_k+i{Q_k \ov g })-{1\ov
x(u_{l}^{(\a)})}} \sqrt{\frac{x(u_k+i{Q_k \ov g })}{x(u_k-i{Q_k
\ov g })}}\, \prod_{M=1}^{\infty}
\prod_{l=1}^{N_{M|vw}^{(\a)}}S_{xv}^{Q_kM}(u_{k},v_{l,M}^{(\a)})\,,~~~~
\eea Here the $\sl(2)$ S-matrix $S_{\sl(2)}^{QQ'}$ in the uniform
light-cone gauge \cite{AFrev} with the gauge parameter $a=0$ can
be written in the form \cite{D2,R,AFb, BJ} \bea\la{Ssl2}
S_{\sl(2)}^{QQ'}(u,u')= S^{QQ'}(u-u')^{-1} \,
\Sigma_{QQ'}(u,u')^{-2}\,, \eea where $S^{QQ'}$ is given by \bea
\la{sqq} S^{QQ'}(u-u')&=&
 \frac{u-u' -{i\ov g}(Q+Q')}{u-u' +{i\ov g}(Q+Q')}  \frac{u-u' -{i\ov g}(Q'-Q)}{u-u' +{i\ov g}(Q'-Q)}\\\nonumber
 &&~~~~~~~~~~~\times\prod_{j=1}^{Q-1}\left(\frac{u-u' -{i\ov g}(Q'-Q+2j)}{u-u' +{i\ov g}(Q'-Q+2j)}\right)^2  \, . ~~~~~~\eea
Here  $\Sigma^{QQ'}(u,u')$ is related to the dressing factor
$\sigma_{QQ'}$ as follows \bea \Sigma_{QQ'}(u,u') =
\sigma_{QQ'}(u,u')\prod_{j=1}^Q\prod_{k=1}^{Q'} {1-{1\ov x(u+{i\ov
g}(Q+2-2j)) x(u'+{i\ov g}(Q'-2k))}\ov 1-{1\ov x(u+{i\ov
g}(Q-2j))x(u'+{i\ov g}(Q'+2-2k))}}\,. \eea
Finally, the auxiliary
S-matrix is given by
\bea\nonumber S_{xv}^{QM}(u,u') &=&
\frac{x(u-i{Q \ov g })-x(u'+i{M \ov g})}{x(u+i{Q \ov g })-x(u'+i{M
\ov g})}\, \frac{x(u-i{Q \ov g })-x(u'-i{M \ov g})}{x(u+i{Q \ov g
})-x(u'-i{M \ov g})}\, \frac{x(u+i{Q \ov g })}{x(u-i{Q \ov g
})}~~~~\\\la{Sxv} &\times
&\prod_{j=1}^{M-1}\frac{u-u'-\frac{i}{g}(Q-M+2j)}{u-u'+\frac{i}{g}(Q-M+2j)}\,.~~~~~~~~~
\eea

Taking the logarithmic derivative of (\ref{BEQ}) with respect to $u_k$, we get in the thermodynamic limit the following integral equation for the densities of $Q$-particles and holes
\begin{eqnarray}\la{tbaQ}
\rho_Q(u) + \br_Q(u) &=& {R\ov 2\pi}{d\tp^Q(u)\ov du} + \sum_{Q'=1}^\infty K_{\sl(2)}^{QQ'}\star\rho_{Q'} \\\nonumber
&&~~~~~~~~+  \sum_{\a=1}^2\left[ K^{Qy}_-\star\rho_{y^-}^{(\a)}+ K^{Qy}_+\star\rho_{y^+}^{(\a)}+\sum_{M'=1}^\infty K^{QM'}_{xv}\star\rho_{M'|vw}^{(\a)}  \right] \,.~~~~
\end{eqnarray}
Here  the kernels $K's$ are
\bea\la{Ks}
K_{\sl(2)}^{QQ'}(u,u')&=&{1\ov 2\pi i}{d\ov du}\log S_{\sl(2)}^{QQ'}(u,u')\,,\\
K^{Qy}_-(u,u')&=&1\ov 2\pi i}{d\ov du}\log \frac{x(u-i{Q\ov g})-x(u')}{x(u+i{Q\ov g})-x(u')} \sqrt{{\frac{x(u+i{Q\ov g})}{x(u-i{Q\ov g})}} \,,\\
K^{Qy}_+(u,u')&=&1\ov 2\pi i}{d\ov du}\log \frac{x(u-i{Q\ov g})-{1\ov x(u')}}{x(u+i{Q\ov g})-{1\ov x(u')}} \sqrt{{\frac{x(u+i{Q\ov g})}{x(u-i{Q\ov g})}} \,,\\
K^{QM'}_{xv}(u,u')&=&{1\ov 2\pi i}{d\ov du}\log
S_{xv}^{QM'}(u,u')\,, \eea where the operation $\star$ is defined
in (\ref{starp}). Some of the kernels can be expressed in terms of
the basic kernels $K_M(u)$ and $K(u,v)$, see appendix
\ref{app:kern}, where also many important properties of the
kernels are listed.


\subsection*{Equations for $y$-particles  densities with ${\rm Im}(y)<0$ }
Next, we take a $y^{(\a)}$-particle with the root
$y^{(\a)}_k=x(u_k^{(\a)})$ and rewrite the equation (3.6) in
\cite{AFsh} in the following form \bea\la{BEy}\hspace{-0.4cm}
-1=\prod_{l=1}^{K^{\mathrm{I}}}\frac{x(u_k^{(\a)})-x^{-}_{l}}{x(u_k^{(\a)})-x^{+}_{l}}\sqrt{\frac{x_l^+}{x_l^-}}
 \prod_{M=1}^{\infty} \prod_{l=1}^{N_{M|vw}^{(\a)}}\frac{u_{k}^{(\a)}-v_{l,M}^{(\a)}-i{M\ov g}}{u_{k}^{(\a)}-v_{l,M}^{(\a)}+i{M\ov g}} \, \prod_{l=1}^{N_{N|w}^{(\a)}}\frac{u_{k}^{(\a)}-w_{l,M}^{(\a)}-i{M\ov g}}{u_{k}^{(\a)}-w_{l,M}^{(\a)}+i{M\ov g}} \,,~~~~~~
\eea where $x_l^\pm = x(u_{l,Q} \pm i{Q\ov g})$ for a $Q$-particle with the real rapidity $u_{l,Q}$.

\smallskip

Taking the logarithmic derivative of (\ref{BEy}) with respect to $u_k^{(\a)}$, we get in the thermodynamic limit  the following integral equation for the densities of $y^-$-particles and holes
\begin{eqnarray}\la{tbaym}
\rho_{y^-}^{(\a)}(u) + \br_{y^-}^{(\a)}(u) = \sum_{M'=1}^\infty\left[ K^{yM'}_-\star\rho_{M'} -K^{yM'}_{vw}\star\left(\rho_{M'|vw}^{(\a)}+\rho_{M'|w}^{(\a)}\right)  \right] \,.~~~~
\end{eqnarray}
Here the kernels $K's$ are
\bea
K^{yM'}_-(u,u')&=&{1\ov 2\pi i}{d\ov du}\log \frac{x(u)-x(u'+i{M'\ov g})}{x(u)-x(u'-i{M'\ov g})}\sqrt{\frac{x(u'-i{M'\ov g})}{x(u'+i{M'\ov g})}}\,,\\
K^{yM'}_{vw}(u,u')&=&K_{M'}(u-u')={1\ov 2\pi i}{d\ov du}\log \frac{u-u' - i{M'\ov g}}{u-u' + i{M'\ov g}}\,.\eea

\subsection*{Equations for $y$-particle densities with ${\rm Im}(y)>0$ }

In the second case ${\rm Im}(y)>0$ the root $y^{(\a)}_k=1/x(u_k^{(\a)})$, and we get in the thermodynamic limit the following integral equation for the densities of $y^+$-particles and holes
\begin{eqnarray}\la{tbayp}
\rho_{y^+}^{(\a)}(u) + \br_{y^+}^{(\a)}(u) = \sum_{M'=1}^\infty\left[ K^{yM'}_+\star\rho_{M'} +K_{M'}\star\left(\rho_{M'|vw}^{(\a)}+\rho_{M'|w}^{(\a)}\right)  \right] \,.~~~~
\end{eqnarray}
Here the kernel $K_{yM'}^+$ is
\bea
K^{yM'}_+(u,u')&=&{1\ov 2\pi i}{d\ov du}\log \frac{{1\ov x(u)}-x(u'-i{M'\ov g})}{{1\ov x(u)}-x(u'+i{M'\ov g})}\sqrt{
\frac{x(u'+i{M'\ov g})}{x(u'-i{M'\ov g})}
}\,.\eea

\subsection*{Equations for $vw$-string  densities}
Then, we take a $M|vw$-string with the coordinates $u^{(\a)}_{k,M}$, and rewrite the equation (3.13) in \cite{AFsh} in the following form
\bea\la{BEvw3}
\hspace{-0.4cm}(-1)^M=\prod_{l=1}^{K^{\mathrm{I}}}S_{xv}^{Q_lM}(u_{l},u_{k,M}^{(\a)})\prod_{l=1}^{N_y^{(\a)}}\frac{u_{k,M}^{(\a)}-v_{l}^{(\a)}-i{M\ov g}}{u_{k,M}^{(\a)}-v_{l}^{(\a)}+i{M\ov g}}
 \prod_{M'=1}^{\infty} \prod_{l=1}^{N_{M'|vw}^{(\a)}}S^{MM'}(u_{k,K}^{(\a)}-v_{l,M'}^{(\a)})\, ,~~~~~~
\eea
where the auxiliary S-matrices are given by (\ref{sqq}) and (\ref{Sxv}).

Taking the logarithmic derivative of (\ref{BEvw3}) with respect to $u_k$, we get in the thermodynamic limit  the following integral equation for the densities of $vw$-strings and holes
\begin{eqnarray}\la{tbavw}
\rho_{M|vw}^{(\a)}(u) + \br_{M|vw}^{(\a)}(u) &=& \sum_{M'=1}^\infty\left[ K^{MM'}_{vwx}\star\rho_{M'}- K^{MM'}_{vv}\star \rho_{M'|vw}^{(\a)}  \right]~~~~ \\\nonumber
&&~~~~~~~~~-K^{My}_{wv}\star\left(\rho_{y^-}^{(\a)} +\rho_{y^+}^{(\a)} \right)\,.~~~~
\end{eqnarray}
Here the kernels $K's$ are
\bea\la{Kvw}
K^{MM'}_{vwx}(u,u')&=&-{1\ov 2\pi i}{d\ov du}\log S_{xv}^{M'M}(u',u)\,,\\\la{Kmmp}
K^{MM'}_{vv}(u,u')&=&K_{MM'}(u-u')={1\ov 2\pi i}{d\ov du}\log S^{MM'}(u-u')\,,\\
K^{My}_{wv}(u,u')&=&K_M(u-u')\,,\eea
and they all are positive.

\subsection*{Equations for $w$-string  densities}
Finally we take a $M|w$-string with the coordinates $u^{(\a)}_{k,M}$, and rewrite the equation (3.9) in \cite{AFsh} in the following form
\begin{eqnarray}\la{BEw}
(-1)^M&=&\prod_{l=1}^{N_y^{(\a)}}\frac{u_{k,M}^{(\a)}-v_{l}^{(\a)}+i{M\ov g}}{u_{k,M}^{(\a)}-v_{l}^{(\a)}-i{M\ov g}}
\prod_ {N=1}^{\infty}
\prod_{l=1}^{N_{N|w}^{(\a)}}S^{KN}(u_{k,M}^{(\a)}-w_{l,N}^{(\a)})
\, .~~~~~
\end{eqnarray}
Taking the logarithmic derivative of (\ref{BEw}) with respect to $u^{(\a)}_{k,M}$, we get in the thermodynamic limit  the following integral equation for the densities of $w$-strings and holes
\begin{eqnarray}\la{tbaw}
\rho_{M|w}^{(\a)}(u) + \br_{M|w}^{(\a)}(u) = K_M\star\left(\rho_{y^-}^{(\a)}
 +\rho_{y^+}^{(\a)} \right) -\sum_{M'=1}^\infty K_{MM'}\star\rho_{M'|w}^{(\a)}  \,.~~~~
\end{eqnarray}

\section{Free energy and equations for pseudo-energies}
Having found the equations for the densities of particles and holes,
 we can proceed with deriving the integral equations delivering the minimum
 of the free energy per unit length for the mirror theory at temperature $T={1\ov
 L}$. The free energy in the mirror theory
 determines the ground state energy of the light-cone $\AdS$ string theory defined on
 the cylinder with the circumference $L$ equal the light-cone momentum $P_+$. In the case we
 are mostly interested in $L=J$, where $J$ is the angular momentum carried by the string rotating about the equator of S$^5$.

\smallskip

To be precise, the light-cone string theory has two different
sectors. The first sector contains even winding number string
states and it has fermions subject to periodic boundary
conditions. This is the sector which has a BPS ground state whose
energy should not receive any quantum corrections. Since the
fermions are periodic, the ground state energy in fact is
determined not by the free energy but by Witten's index  of the
mirror theory. The second sector has anti-periodic fermions and
has a non-BPS ground state whose energy is determined by the
mirror free energy. To describe both sectors in one go, we
consider a generalized free energy \cite{CIV} defined by the following
equation
\bea\la{feg} \F_\g(L)= {\cal E}  - {1\ov L} S + {i\g\ov
L}( N_F^{(1)}- N_F^{(2)})\,, \eea where ${\cal E}$ is the energy
per unit length carried by $Q$-particle densities \bea\la{cale}
{\cal E}=\int {\rm d}u \sum_{Q=1}^\infty \tH^Q(u)\rho_Q(u)\, .
\eea Here  $\tH^Q(u)$ is a $Q$-particle energy defined in
(\ref{Hu}), and $S$ is the total entropy,
 \bea\nonumber
S = \int {\rm d}u\,\left[ \sum_{Q=1}^\infty \cs\left(\rho_Q \right)+ \sum_{\a=1}^2\left(
 \cs\left(\rho_{y^-}^{(\a)} \right)+\cs\left(\rho_{y^+}^{(\a)} \right) + \sum_{M=1}^\infty\left( \cs\big(\rho_{M|vw}^{(\a)} \big)+\cs\big(\rho_{M|w}^{(\a)} \big) \right) \right)\right]\,,
\eea
 where $\cs(\rho)$ denotes the entropy function of densities of particles and holes
\bea
\cs(\rho)= \rho\log\left(1+{\br\ov\rho}\right) + \br\log\left(1+{\rho\ov\br}\right)\,.
\eea
 Then, $1/L$ is the temperature of the mirror theory,
$i\g/L$ plays the role of a chemical potential, and
$N_F^{(\a)}$ is the fermion number which counts the number of  $y^{(\a)}$-particles which are the only fermions in our system
\bea\la{nf12}
&&  N_F^{(1)}- N_F^{(2)}=\int {\rm d}u\,  ( \rho_{y^-}^{(1)}(u)+ \rho_{y^+}^{(1)}(u) -  \rho_{y^-}^{(2)}(u)- \rho_{y^+}^{(2)}(u))\,.
\eea

The relative minus sign between $N_F^{(1)}$ and  $N_F^{(2)}$ is
needed for the reality of the free energy, for relativistic examples see \cite{CIV}. In principle, the
choice of ``$+$'' or ``$-$'' sign should not matter at $\g=\pi$
where (\ref{feg}) becomes Witten's index. If $\g=0$ we get the
usual free energy.

\subsection{Derivation of the equations}

Thus, we need to minimize the free energy at temperature $T=1/L$ defined by the following equation
\bea\la{fre}
&&\F_\g(L)=\int {\rm d}u\,\left[ \sum_{Q=1}^\infty \tH^Q(u)\rho_Q(u) - {i\g\ov L}\, \sum_{\a=1}^2(-1)^\a( \rho_{y^-}^{(\a)}(u)+ \rho_{y^+}^{(\a)}(u))  - {S\ov L} \right]\,.~~~
\eea
The consideration is standard and general. We first write the free energy  in compact form as follows
\bea\la{frec}
\F_\g(L)
=\int {\rm d}u\, \sum_{k}\left[ \tH_{k}\, \rho_k - {i\g_k\ov L}\, \rho_k  - {1\ov L}\cs(\rho_k) \right]\,,
\eea
where $\tH_{k}$ and $\g_k$ do not vanish only for $Q$- and $y$-particles, respectively.

\smallskip

Since the densities of particles and holes satisfy the Bethe equations (\ref{tbacom}), their variations are not independent but are subject to
\begin{eqnarray}\la{tbacomv}
\delta\rho_k(u) + \delta\br_k(u) =K_{kj}\star\delta\rho_j \,.~~~~
\end{eqnarray}
Here and in what follows the summation over the repeated indices is assumed.
Expressing the variations of the densities of holes in terms of the variations of the densities of particles, one finds the variation of the entropy function
\bea\la{entropyv}
\delta \cs(\rho_k) =(\e_k-i\g_k)\delta\rho_i + \log\left(1+e^{i\g_k-\e_k} \right)K_{kj}\star\delta \rho_j\,,
\eea
where the pseudo-energies $\e_k$ are defined through
\bea
e^{i\g_k-\e_k} = {\rho_k\ov\br_k}\,.
\eea
Then,  using the extremum condition $\delta\F_\g(L)=0$, one derives the following set of TBA equations
\bea\la{TBAeq}
\e_k = L\, \tH_{k} - \log\left(1+e^{i\g_j-\e_j} \right)\star K_{jk}\,,
\eea
where the right action of the kernels $K_{jk}$ defined by (\ref{starp2}) is used.

\smallskip

Note also that $\e_{y^\pm}$ is defined only for $|u|\le 2$. In
principle, one  could extend it to $|u|>2$ by saying that
$e^{-\e_{y^\pm}}=0$ for $|u|>2$. It might, however, lead to
discontinuities of the $y$-particles pseudo-energies.

\smallskip

Taking into account that the entropy function can be written in the form
\bea\la{entropy2}
\cs(\rho_k) ={R\ov 2\pi}{d\tp_k\ov du}\log\left(1+e^{i\g_k-\e_k} \right)
 + (\e_k-i\g_k)\rho_k + \log\left(1+e^{i\g_k-\e_k} \right)K_{kj}\star \rho_j\,,~~~~~
\eea one finds that at the extremum (\ref{TBAeq}) the free energy
is equal to \bea \la{frec2} \F_\g(L) =-{R\ov L}\int {\rm d}u\,
\sum_{k}{1\ov 2\pi}{d\tp_k\ov du}\log\left(1+e^{i\g_k-\e_k}
\right)\,. \eea Finally, one uses that  the energy of the ground
state of the light-cone string theory is related to the free
energy of the mirror model as
\bea E_\g(L)= \lim_{R\to\infty}{L\ov
R}\F_\g(L)\,,
 \eea
 and gets the following expression
 \bea
\la{energyL} E_\g(L) &=&-\int {\rm d}u\, \sum_{Q=1}^\infty{1\ov
2\pi}{d\tp^Q\ov du}\log\left(1+e^{-\e_Q} \right)\,. \eea
We see
that  for $\g=\pi$  the necessary condition for the ground state
energy to vanish for any $L$ is \bea\la{solnec} e^{-\e_{Q}}=0\
~~{\rm for\ any\ }~~\ Q\,. \eea It also imposes restrictions on
pseudo-energies of other string configurations.

\subsection{TBA equations explicitly}
Here we list all TBA equations (\ref{TBAeq}) explicitly taking into account that
\bea
\g_Q=\g_{M|vw}^{(\a)}=\g_{M|w}^{(\a)}=0\,,\quad \g_{y^\pm}^{(\a)}=(-1)^\a\pi +h_\a\,,\quad h_\a=(-1)^\a h\,,
\eea
and we used this representation for $\g_{y^\pm}^{(\a)}$ to handle more efficiently the physically more interesting case with $\g=\pi$.

Assuming summation over repeated indices and the index $\a$ in the equation for $Q$-particles, the TBA equations for the pseudo-energies take the following form
\begin{itemize}
\item $Q$-particles
\bea\la{TbaQ}
&&\e_Q = L\, \tH_{Q} - \log\left(1+e^{-\e_{Q'}} \right)\star K_{\sl(2)}^{Q'Q} - \log\left(1+e^{-\e_{M'|vw}^{(\a)}} \right)\star  K^{M'Q}_{vwx} \\\nonumber
 &&~~~~~~~~~~~~~~ - \log\left(1-e^{ih_\a-\e_{y^-}^{(\a)}} \right)\star K^{yQ}_- - \log\left(1-e^{ih_\a-\e_{y^+}^{(\a)}} \right)\star K^{yQ}_+\,.~~~~~~
\eea

\item $y$-particles
\bea\la{Tbay}
&&\e_{y^\pm}^{(\a)} =  - \log\left(1+e^{-\e_{Q}} \right)\star K^{Qy}_\pm +\log {1+e^{-\e_{M|vw}^{(\a)}} \ov1+e^{-\e_{M|w}^{(\a)}}}\star K_{M}\,.~~~~~~
\eea

\item $M|vw$-strings
\bea\la{Tbavw}
&&\e_{M|vw}^{(\a)} =  - \log\left(1+e^{-\e_{Q'}} \right)\star K^{Q'M}_{xv}\\\nonumber
 &&~~~~~~~~~~~ +  \log\left(1+e^{-\e_{M'|vw}^{(\a)}} \right)\star K_{M'M} - \log{1-e^{ih_\a -\e_{y^+}^{(\a)}}\ov 1-e^{ih_\a -\e_{y^-}^{(\a)}} }\star K_M\,.
\eea

\item $M|w$-strings
\bea\la{Tbaw}
\e_{M|w}^{(\a)} =  \log\left(1+e^{-\e_{M'|w}^{(\a)}} \right)\star K_{M'M}
- \log{1-e^{ih_\a -\e_{y^+}^{(\a)}}\ov 1-e^{ih_\a -\e_{y^-}^{(\a)}} }\star K_M\,.
\eea

\end{itemize}

We  see that  (\ref{solnec})  solves these equations for $h=0$ if
the pseudo-energies of $y$-particles satisfy $e^{-\e_y}=1$. A proper way to analyze these system for $h=0$ is to consider the perturbation theory  in small $h$.
A rough estimate seems to show that $\e_{y^-}^{(\a)} \sim h$ and $e^{-\e_{Q}}\sim h^2$.
We postpone detailed analysis for future.

\section{Simplifying the TBA equations}

In this section we simplify the system of TBA equations
(\ref{TbaQ}-\ref{Tbaw}) by reducing most of the equations to a
local form. The local form can be also readily used to derive
equations for Y-functions which coincide with (or are inverse to)
exponentials of pseudo-energies. We will see that the recently
conjecture Y-system \cite{GKV2} holds only for values of the
spectral parameter $u$ satisfying the inequality $|u|<2$. For
other values of $u$ the TBA equations for $Q$-particles cannot be
apparently reduced to the Y-type equations.

\smallskip

We introduce the Y-functions as \bea\la{Yf} Y_{Q} =
e^{-\e_{Q}}\,,\quad Y_{M|vw}^{(\a)} = e^{\e_{M|vw}^{(\a)}}\,,\quad
Y_{M|w}^{(\a)} = e^{\e_{M|w}^{(\a)}}\,,\quad Y_\pm^{(\a)}
=e^{\e_{y^\pm}^{(\a)}}\,, \eea and use the following universal
kernel \bea\la{invK0} \left( K  + 1\right)_{MN}^{-1} = \delta_{MN}
- s\left( \delta_{M+1,N}+ \delta_{M-1,N}\right)\,,\quad s(u)=
{g\ov 4\cosh {g\pi u\ov 2}}\,,~~~~~~ \eea that is inverse to the
kernel $ K_{NQ}  + \delta_{NQ}$ \bea \la{invKK}
\sum_{N=1}^\infty\left( K + 1\right)_{MN}^{-1} \left( K_{NQ}  +
\delta_{NQ}\right)= \delta_{MQ}\,.~~~~~ \eea For more properties
of the inverse kernel see appendix \ref{app:kern}.

\subsection*{TBA and Y-equations for  $w$-strings}
We begin our consideration with the simplest case of $w$-strings.
We apply the inverse kernel (\ref{invK0}) to (\ref{Tbaw}), and get the following equation
\bea\la{Yforw}
\log Y_{M|w}^{(\a)}= I_{MN} \log(1 +  Y_{N|w}^{(\a)} )\star s
+\delta_{M1}\, \log{1-{e^{ih_\a}\ov Y_-^{(\a)}}\ov 1-{e^{ih_\a}\ov Y_+^{(\a)}} }\star s\,,~~~~~
\eea
where $I_{MN}$ is the incidence matrix
\bea\la{incm}
I_{MN} =  \delta_{M+1,N}+ \delta_{M-1,N}\,,
\eea
and we used the following identity
\bea\la{kmnikm}
\sum_{N=1}^\infty\left( K + 1\right)_{MN}^{-1} \,K_{N} =s\,\delta_{M1}\,.
\eea
Since the functions
$Y_{\pm}^{\a}$ are defined on the interval $-2<u<2$, the 
integral in the last term of eq.(\ref{Yforw}) is taken from $-2$ to $2$.

\smallskip

To derive the Y-equations for $w$-strings, it is convenient to
define the operator $s^{-1}$ that acts on functions of the
rapidity variable $u$ as follows \bea\la{defs} (f\star s^{-1})(u)
= \lim_{\e\to 0^+} \big[ f(u+{i\ov g} - i\e )+ f(u-{i\ov g} + i\e
)\big]\,. \eea It satisfies the obvious identity \bea\nonumber
(s\star s^{-1})(u) =\delta(u)\,. \eea The operator $s^{-1}$ has
however a large null space, and as a result in general
$$
f\star s^{-1}\star s \neq f\,.
$$
This also means that one may loose information by acting by  the operator $s^{-1}$ on an equation. We will see examples of such a loss in what follows.

\smallskip

By applying the $s^{-1}$ operator to both sides of the equation one immediately gets the following Y-equations for $w$-strings
\bea\la{YwM}
Y_{M|w}^{(\a)+}\,Y_{M|w}^{(\a)-} &=& \left( 1+Y_{M-1|w}^{(\a)} \right)\left( 1+Y_{M+1|w}^{(\a)} \right)  \quad {\rm if}\ \ M\ge 2\,,
\\ \la{Yw1}
Y_{1|w}^{(\a)+}\,Y_{1|w}^{(\a)-} &=& \left( 1+Y_{2|w}^{(\a)} \right){1-{e^{ih_\a}\ov Y_-^{(\a)}}\ov 1-{e^{ih_\a}\ov Y_+^{(\a)}} } \,, 
\qquad |u|\leq 2 \, ,\\  \la{Yw1b}
Y_{1|w}^{(\a)+}\,Y_{1|w}^{(\a)-} &=& 1+Y_{2|w}^{(\a)}\, , \qquad\qquad\qquad\quad  |u|>2\, ,
\eea
where we introduce the notation $Y_{M|w}^{(\a)\pm}(u)\equiv Y_{M|w}^{(\a)}(u\pm {i\ov g}\mp i0)$.

\smallskip

These formulae show that the form of the 
Y-equations for $M=1$ is not uniform with respect to the parameter $u$. 
The reason behind is that $Y_{\pm}^{(\a)}$ are supported on the interval
$(-2,2)$. Only eqs.(\ref{YwM}) and (\ref{Yw1}) have appeared in \cite{GKV1} and they were assumed to hold for all values of $u$.
Obviously, the uniform expression could be achieved provided $Y_{\pm}^{(\a)}$ admit such an analytic continuation to the complex $u$-plane 
that $Y_{+}^{(\a)}(u)=Y_{-}^{(\a)}(u)$ for $|u|>2$. This continuation should be, however, compatible with the whole set of TBA system.
Currently it is unclear if this is indeed the case.

\subsection*{TBA and Y-equations for  $y$-particles}
Next we consider $w$-strings.
Equations (\ref{Tbay}) for the pseudo-energies of $y$-particles can be written in the form
\bea\la{Ttbaym}
&&\log Y_-^{(\a)} =-{1\ov 2}\log\left(1+Y_Q \right)\star \left(  K_{Qy} + K_Q\right)+\log {1+{1\ov Y_{M|vw}^{(\a)}}
 \ov1+{1\ov Y_{M|w}^{(\a)}}}\star K_M \, ,\\
\la{Ttbayp}
&&\log Y_+^{(\a)}  ={1\ov 2}\log\left(1+Y_Q \right)\star \left(  K_{Qy} - K_Q\right)+\log {1+{1\ov Y_{M|vw}^{(\a)}} \ov1+{1\ov Y_{M|w}^{(\a)}}}\star K_M\,,
\eea
where we used that the kernels $K_\pm^{Qy}$ can be expressed in terms of $K_Q$ and  the kernel $K_{Qy}$ defined in (\ref{KQyuv}).

\smallskip

By adding and subtracting, these equations can be cast in the form
\bea\la{Yfory1}
&&\log {Y_+^{(\a)} \ov Y_-^{(\a)}}=  \log(1 +  Y_{Q})\star K_{Qy} \,,\\
\la{Yfory2} &&\log {Y_+^{(\a)}  Y_-^{(\a)}} =  - \log\left(1+Y_Q
\right)\star K_Q+2\log {1+{1\ov Y_{M|vw}^{(\a)}}
 \ov1+{1\ov Y_{M|w}^{(\a)}}}\star K_M\,. \eea The last term in
(\ref{Yfory2}) can be replaced by (\ref{Yforvww}) or
(\ref{Yforvww2}) reducing all non-local terms to the ones
depending on Y-functions of $Q$-particles only.

\smallskip

To derive Y-equations for $y$-particles, we need  the following
identities \bea \left(  K_{Qy} + K_Q\right)\star s^{-1}=2
K_{xv}^{Q1} + 2\delta_{Q1}\,,\quad K_M\star s^{-1} = K_{M1} +
\delta_{M1}\,. \eea  We get with their help \bea\la{tbaym2} \log
Y_-^{(\a)} \star s^{-1} &=&-\log\left(1+Y_1
\right)-\log\left(1+Y_Q \right)\star K_{xv}^{Q1} \\\nonumber
&+&\log {1+{1\ov Y_{1|vw}^{(\a)}} \ov1+{1\ov Y_{1|w}^{(\a)}}}+\log
{1+{1\ov Y_{M|vw}^{(\a)}} \ov1+{1\ov Y_{M|w}^{(\a)}}}\star K_{M1}
\,. \eea Now we subtract (\ref{Tbaw}) from (\ref{Tbavw}) for
$M=1$, and obtain \bea\la{vwmw} \log {Y_{1|vw}^{(\a)} \ov
Y_{1|w}^{(\a)}} = -\log\left(1+Y_Q \right)\star K_{xv}^{Q1} + \log
{1+{1\ov Y_{M|vw}^{(\a)}} \ov1+{1\ov Y_{M|w}^{(\a)}}}\star
K_{M1}\, . \eea Finally, subtracting (\ref{vwmw}) from
(\ref{tbaym2}), we derive the following Y-equation for
$y_-$-particles \bea\la{tbaym4} Y_{-}^{(\a)+}\,Y_{-}^{(\a)-}  &=&
{1+ Y_{1|vw}^{(\a)} \ov1+ Y_{1|w}^{(\a)}} {1\ov 1+Y_1 } \,. \eea
Repeating the same procedure with $Y_{+}^{(\a)} $, we get \bea
\la{tbayp2} \log Y_{+}^{(\a)+}\,Y_{+}^{(\a)-} &=&\log\left(1+Y_Q
\right)\star (2K_{xv}^{Q1} -  K_{Q1}) +\log {1+ Y_{1|vw}^{(\a)}
\ov1+ Y_{1|w}^{(\a)}} \\\nonumber &=&2 \log\left(1+Y_Q
\right)\star {\cal K}_{2}^{Q1}  +\log {1+ Y_{1|vw}^{(\a)} \ov1+
Y_{1|w}^{(\a)}}
 \,,
\eea
where the kernel $ {\cal K}_{2}^{QQ'} $ is defined in (\ref{kxvNN4}). This equation cannot be reduced to the usual local Y-system form.

\smallskip

We stress again that the equations for $Y_{\pm}^{(\a)}$ are valid for $u$ being in the interval $(-2,2)$ and the analytic continuation 
for $|u|>2$, we have discussed after eq.(\ref{Yw1b}), would impose a consistency condition on the functions $Y_Q$ due to eq.(\ref{Yfory1}).

\subsection*{TBA and Y-equations for  $vw$-strings}
Now we can discuss $vw$-strings. We apply the inverse kernel
(\ref{invK0}) to (\ref{Tbavw}), use an identity \bea\la{eqq3}
K^{QQ''}_{xv}\star \left( K +1\right)_{Q''Q'} ^{-1} &=&
 \delta_{Q-1,Q'}s +\delta_{Q'1}  K_{Qy}\star s \,,~~~~~
\eea and, as a result, obtain the following equation
\bea\la{Yforvw} \log Y_{M|vw}^{(\a)}&=&I_{MN} \log(1 +
Y_{N|vw}^{(\a)} )\star s
+\delta_{M1}\,   \log{1-{e^{ih_\a} \ov Y_{-}^{(\a)}}\ov 1-{e^{ih_\a} \ov Y_{+}^{(\a)}} }\star s\\\nonumber
&-& \log(1 +  Y_{M+1}) \star s -\delta_{M1} \log(1 +  Y_{Q})\star K_{Qy} \star s\,,~~~~~
\eea
where in $K_{Qy}\star  s$ we integrate over $[-2,2]$.
By using (\ref{Yfory1}), we can rewrite (\ref{Yforvw})  in the final local form
\bea\la{Yforvw2}
\hspace{-0.3cm}\log Y_{M|vw}^{(\a)}=\Big( I_{MN} \log(1 +  Y_{N|vw}^{(\a)} )
+\delta_{M1}  \log{1-e^{-ih_\a}Y_-^{(\a)}\ov 1-e^{-ih_\a}Y_+^{(\a)}}- \log(1 +  Y_{M+1})\Big)\star s\,,~~~~~
\eea
Now applying the $s^{-1}$ operator to both sides of the equation, one gets the following Y-equations for $vw$-strings
\bea\la{YvwM}
Y_{M|vw}^{(\a)+}\, Y_{M|vw}^{(\a)-} &=& \left( 1+Y_{M-1|vw}^{(\a)} \right)\left( 1+Y_{M+1|vw}^{(\a)} \right)\, {1\ov 1+Y_{M+1}}  \quad {\rm if}\ 
\ M\ge 2\,,\\\la{Yvw1}
Y_{1|vw}^{(\a)+}\, Y_{1|vw}^{(\a)-} &=& { 1+Y_{2|vw}^{(\a)} \ov 1+Y_{2}} \, {1-e^{-ih_\a}Y_-^{(\a)}\ov 1-e^{-ih_\a}Y_+^{(\a)}} \,, \qquad |u|\leq 2\, , \\
Y_{1|vw}^{(\a)+}\, Y_{1|vw}^{(\a)-} &=& { 1+Y_{2|vw}^{(\a)} \ov 1+Y_{2}}  \,, \qquad\qquad\qquad\qquad~ |u|>2\, .
\eea

Finally, subtracting (\ref{Yforw}) from  (\ref{Yforvw}), we derive
the following equation \bea\la{Yforvww} \log{1 +  {1\ov
Y_{M|vw}^{(\a)}} \ov 1 +  {1\ov Y_{M|w}^{(\a)}} } \star K_M&=&
\log{1 +  Y_{1|vw}^{(\a)} \ov 1 +  Y_{1|w}^{(\a)} }  \star
s\\\nonumber
 &+& \log(1 +  Y_{M+1}) \star s\star K_M+  \log(1 +  Y_{M})\star K_{My} \star s\star K_1\,,~~~~~
\eea
and by using (\ref{Yfory1}), one can rewrite (\ref{Yforvww}) in the form
\bea\la{Yforvww2}
\log{1 +  {1\ov Y_{M|vw}^{(\a)}} \ov 1 +  {1\ov Y_{M|w}^{(\a)}} }\star K_M &=&  \log{1 +  Y_{1|vw}^{(\a)} \ov 1 +  Y_{1|w}^{(\a)} }  \star s\\\nonumber
 &+& \log(1 +  Y_{M+1}) \star s\star K_M+  \log{Y_+^{(\a)}\ov Y_-^{(\a)}} \star s\star K_1\,.~~~~~
\eea
These equations can be used to simplify eq.(\ref{Yfory2}) for $y$-particles.

\subsection*{TBA equations for  $Q$-particles }
At last,  we  discuss the most complicated case of $Q$-particles.
Our analysis is only partially complete here because of a lack of
understanding of the properties of the dressing kernel in the
mirror model. Still, we will be able to demonstrate that the
transition from the TBA equations to Y-equations is only possible
for $u$ taking values in the interval from $-2$ to 2.

\smallskip

Equations for the pseudo-energies of $Q$-particles can be written in the form
\bea\nonumber
-\log Y_Q &=& L\, \tH_{Q} +\log\left(1+Y_{Q'} \right)\star ( K_{Q'Q} + 2 K^\Sigma_{Q'Q}) -  \log\left(1+{1\ov Y_{M|vw}^{(\a)}} \right)\star K^{MQ}_{vwx}~~~~\\\la{YforQ}
 & -&{1\ov 2}  \log{1-{e^{ih_\a} \ov Y_{-}^{(\a)}}\ov 1-{e^{ih_\a} \ov Y_{+}^{(\a)}} }\star K_Q - {1\ov 2} \log\left(1-{e^{ih_\a} \ov Y_{-}^{(\a)}} \right)\left(1-{e^{ih_\a} \ov Y_{+}^{(\a)}}\right)\star K_{yQ}\,,
\eea
where the summation over $\a$ is assumed, and  we use (\ref{Ssl2}), (\ref{Ks}) and (\ref{Kmmp}) to represent the $Q$-particles kernel in the following form
\bea\la{Kqqsl2}
K_{\sl(2)}^{QQ'}(u,u')&=& - K_{QQ'}(u-u') - 2 K^\Sigma_{QQ'}(u,u')\,,
\eea
where we introduce the dressing phase kernel
\bea
\la{KSig}
 K^\Sigma_{QQ'}(u,u') &=& {1\ov 2\pi i}{d\ov du}\log \Sigma_{QQ'}(u,u')\,.
\eea We now should apply the inverse kernel (\ref{invK0}) to
(\ref{YforQ}). To this end, in addition to  formula
(\ref{kmnikm}), we need the following formulae \bea\la{EQKi}
 \tH_{Q'}\star (K + 1)^{-1}_{Q'Q} &=&\delta_{Q1}( \tH_{1} - \tH_{2}\star s) = \delta_{Q1} \,\check{\cal E}\star s\, .~~~~~~~~~
\eea Here
 \bea
(\check{\cal E}\star s)(v) &=&   \int_{-\infty}^{\infty}{\rm d}u'\, \check{\cal E}(u')s(u'-v)\\
\nonumber &=& 2\Big( \int_{-\infty}^{-2} -  \int^{\infty}_{2}
\Big){\rm d}u'\, \log \left(\frac{1}{2}
   \left|u'-\sqrt{u'^2
   -4}\, \right|\right)s(u'-v)\,,
\eea
where we introduce the function
\bea\la{cEu}
\check{\cal E}(u)= 2\log \left(\frac{1}{2}
   \left|u-\sqrt{u^2
   -4}\, \right|\right)\left(\theta(-u-2) -\theta(u-2)\right)\,,
\eea
and $\theta( u)$ is the standard unit step function.

\smallskip

Next, we find the following action of the inverse kernel on
$K_{yQ'}$ \bea\la{KyQKi}
 K_{yQ'}\star (K + 1)^{-1}_{Q'Q} &=&\delta_{Q1}(K_{y1}- K_{y2}\star s) = \delta_{Q1}(s + 2 \check{K}\star s)\,,~~~~~~~~~\\\nonumber
(\check{K}\star s)(u,v) &=&   \int_{-\infty}^{\infty}{\rm d}u'\, \check{K}(u,u')s(u'-v)=\Big( \int_{-\infty}^{-2} -  \int^{\infty}_{2} \Big){\rm d}u'\, \bar{K}(u,u')s(u'-v)\\
\nonumber
&=&
 \pm  \Big( \int_{-\infty}^{-2} +  \int^{\infty}_{2} \Big){\rm d}u'\, K(u,u'\pm i\e)s(u'-v)\,,
\eea where we use the kernels \bea\la{cKuv} \check{K}(u,v)=
\bar{K}(u,v)\left(\theta(-v-2) -\theta(v-2)\right)\,,\quad
\bar{K}(u,v)= \frac{ 1}{2 \pi }\,\frac{
\sqrt{v^2-4}}{\sqrt{4-u^2}}\, {1\ov u-v}   \,.  ~~~~~~~ \eea We
also need similar formulae for $K^{MQ'}_{vwx}$ \bea\nonumber
&&\hspace{-.5cm} K^{MQ'}_{vwx}\star (K + 1)^{-1}_{Q'Q}
=\delta_{M+1,Q}s +\delta_{1Q}(K_{vwx}^{M1}-K_{vwx}^{M2}\star
s)=\delta_{M+1,Q}s +\delta_{1Q}\check{K}_{M}\star s\,,\\\la{KvwKi}
&&\hspace{-.5cm} (\check{K}_{M}\star s)(u,v)=
\int_{-\infty}^{\infty}{\rm d}u'\,
\check{K}_M(u,u')s(u'-v)\\\nonumber &&~~~~~~~~~=  \Big(
\int_{-\infty}^{-2} -  \int^{\infty}_{2} \Big){\rm d}u'\,
(\bar{K}(u+{i\ov g}M,u') + \bar{K}(u-{i\ov g}M,u'))s(u'-v)\\
\nonumber &&~~~~~~~~~ =\pm  \Big( \int_{-\infty}^{-2} +
\int^{\infty}_{2} \Big){\rm d}u'\, (K(u+{i\ov g}M,u'\pm
i\e)+K(u-{i\ov g}M,u'\pm i\e)) s(u'-v)\,, \eea where
\bea\la{cKMuv} \check{K}_M (u,v)= (\bar{K}(u+{i\ov g}M,v) +
\bar{K}(u-{i\ov g}M,v))\left(\theta(-v-2) -\theta(v-2)\right)\,.
\eea The kernels $\check{K}(u,v)$ and $\check{K}_M(u,v)$ obviously
vanish only for $|v|<2$. For other values of $v$ they are
non-trivial and because of that the Y-system would not hold for
$|v|>2$.

\smallskip

The only formula we are missing is the one which gives the action
of the inverse kernel on the dressing phase kernel:
$K^\Sigma_{Q'Q''}\star (K + 1)^{-1}_{Q''Q} = ?$. Unfortunately,
since the structure of the dressing phase in the mirror model is
unknown, we cannot proceed. We believe that the result will be
similar to what we found above for the other kernels \bea\la{KSKi}
K^\Sigma_{Q'Q''}\star (K + 1)^{-1}_{Q''Q}
\stackrel{?}{=}\delta_{1Q}\check{K}_{Q'}^\Sigma\star s\,, \eea
where the kernel $\check{K}_{Q'}^\Sigma(u,v)$ would vanish for
$|v|<2$. Finding a formula for the kernel $\check{K}_{Q'}^\Sigma$
is necessary for understanding the structure of the TBA equations,
 and we hope to address the problem in a future publication.

\smallskip

Applying now the inverse kernel (\ref{invK0}) to (\ref{YforQ}), and assuming that $Q\ge 2$, we get the following equation
\bea\nonumber
-\log Y_{Q}+ \log(Y_{Q-1}Y_{Q+1})\star s&=&\log{(1 +  Y_{Q-1} )(1 +  Y_{Q+1} ) \ov \left(1 +  {1\ov Y_{Q-1|vw}^{(1)}} \right)\left(1 +  {1\ov Y_{Q-1|vw}^{(2)}} \right)}\star  s\\\la{YforQ2}
&+&2\log\left(1+Y_{Q'} \right)\star K^\Sigma_{Q'Q''}\star (K + 1)^{-1}_{Q''Q} \,,~~~~~
\eea
We see that if the formula (\ref{KSKi}) would hold then
applying the $s^{-1}$ operator to both sides of the equation (\ref{YforQ2}),  one gets the following Y-equation for $Q$-particles
\bea\la{YQQ}
{Y_{Q}^{+}\, Y_{Q}^{-}\ov Y_{Q-1}Y_{Q+1}} &=& {\left(1 +  {1\ov Y_{Q-1|vw}^{(1)}} \right)\left(1 +  {1\ov Y_{Q-1|vw}^{(2)}} \right)\ov \left( 1+Y_{Q-1} \right)\left( 1+Y_{Q+1} \right) } \quad {\rm if}\ \ Q\ge 2\,.
\eea

This Y-equation agrees with the corresponding equation of the Y-system in \cite{GKV1} if one identifies $Y_Q=Y_{Q,0}\,,\ Y_{Q|vw}^{(1)} = 1/Y_{Q+1,1} \,,\ Y_{Q|vw}^{(2)} = 1/Y_{Q+1,-1}$.

\medskip

Finally, the equation for $Q=1$ takes the form
\bea
\la{YforQ1}
-\log Y_{1}&=& \log(1 +  {1\ov Y_{2}} )\star s
-\log\left(1-{e^{ih_1} \ov Y_{-}^{(1)}} \right)\left(1-{e^{ih_2} \ov Y_{-}^{(2)}}\right)\star s \\\nonumber
&+&  L\, \check{\cal E}\star s -\log\left(1-{e^{ih_1} \ov Y_{-}^{(1)}} \right)\left(1-{e^{ih_2} \ov Y_{-}^{(2)}}\right)
\left(1-{e^{ih_1} \ov Y_{+}^{(1)}} \right)\left(1-{e^{ih_2} \ov Y_{+}^{(2)}}\right)
\star \check{K}\star s
\\
\nonumber&-& \log\left(1 +  {1\ov Y_{M|vw}^{(1)}} \right)\left(1 +  {1\ov Y_{M|vw}^{(2)}} \right)\star \check{K}_M\star s
\\\nonumber
&+&2\log\left(1+Y_{Q} \right)\star K^\Sigma_{QQ'}\star (K +
1)^{-1}_{Q'1}  \,.~~~~~ \eea We recall that the functions
$Y_{\pm}^{\a}$ are defined on the interval $-2<u<2$ and,
therefore, in all the formulae where they appear the corresponding
integrals are taken from $-2$ to $2$. The equation above clearly
demonstrates the absence of symmetry between $y^-$- and
$y^+$-particles. It does not immediately lead to the corresponding
equation in the Y-system of \cite{GKV1} for arbitrary values of
the rapidity variable $u$ which is the argument of the function
$Y_1$ in (\ref{YforQ1}).

\smallskip

Nevertheless,  applying the $s^{-1}$ operator to both sides of
(\ref{YforQ1}), one gets the following equation \bea\la{YQ0}
{Y_{1}^{+}\, Y_{1}^{-}\ov Y_{2}} &=& {\left(1 -  {e^{ih_1}\ov
Y_{-}^{(1)}} \right)\left(1 -  {e^{ih_2}\ov Y_{-}^{(2)}}
\right)\ov 1+Y_2 }\,  e^{-\Delta}\,,\eea where \bea\la{addit}
\Delta&=&\log\left(1-{e^{ih_1} \ov Y_{-}^{(1)}}
\right)\left(1-{e^{ih_2} \ov Y_{-}^{(2)}}\right) \big(\theta(-u-2)
+\theta(u-2)\big)
\\\nonumber
&+& L\, \check{\cal E} -\log\left(1-{e^{ih_1} \ov Y_{-}^{(1)}}
\right)\left(1-{e^{ih_2} \ov Y_{-}^{(2)}}\right) \left(1-{e^{ih_1}
\ov Y_{+}^{(1)}} \right)\left(1-{e^{ih_2} \ov Y_{+}^{(2)}}\right)
\star \check{K}
\\
\nonumber&-& \log\left(1 +  {1\ov Y_{M|vw}^{(1)}} \right)\left(1 +
{1\ov Y_{M|vw}^{(2)}} \right)\star \check{K}_M +2\log\left(1+Y_{Q}
\right)\star {\check K}^\Sigma_{Q}  \,.~~~~~ \eea Here the first
term on the right hand side guarantees that the second  term on
the right hand side of eq.(\ref{YforQ1}) contributes only for
$|u|<2$.

\smallskip

Since the extra contribution $\Delta$ vanishes for $|u|<2$, in
this case one recovers the
 Y-equation for $Q=1$-particles
\bea\la{YQ1} {Y_{1}^{+}\, Y_{1}^{-}\ov Y_{2}} &=& {\left(1 -
{e^{ih_1}\ov Y_{-}^{(1)}} \right)\left(1 -  {e^{ih_2}\ov
Y_{-}^{(2)}} \right)\ov 1+Y_2 } \quad {\rm if}\ \  |u|<2\,.\eea
This equation also agrees with the Y-system of \cite{GKV1} under
the identification $  {e^{-ih_1} Y_{-}^{(1)}}  =- 1/Y_{1,1} \,,\
{e^{-ih_2} Y_{-}^{(2)}}  = -1/Y_{1,-1} $. Let us stress again that
to derive (\ref{YQ1}) one uses that $\check{\cal E}(u)\,,\
\check{K}(u',u)\,,\   \check{K}_M(u',u)\,,\
\check{K}_{Q}^\Sigma(u',u)$ vanish for $|u|<2$, and assumes the
validity of formula (\ref{KSKi}) for any $Q'$.

\smallskip

Taking into account the important role the Y-system played in the
recent studies of the O(4) chiral model \cite{GKV}, it might seem
reasonable to assume that
 only solutions of the TBA equations  satisfying the extra condition
\bea\la{extrac} \Delta = 2\pi i n\qquad {\rm for\ all}\ \
u\,~~{\rm and~~some}~~n\in\mathbb{Z}, \eea encode the spectrum of
strings on $\ads$, and are relevant for the AdS/CFT
correspondence. Currently, however, we suspend the claim that it
is really the case.  In fact, it seems  that the solution of the
TBA equations leading to the vanishing ground state energy does
not obey the condition, and if it is really the case then the
Y-system is valid only for $|u|<2$. We hope to return to this
question in a future publication.

\smallskip

Let us finally  mention that identifying
 \bea\nonumber
 &&Y_Q=Y_{Q,0}\,,\  {e^{-ih_1} Y_{-}^{(1)}}  =- 1/Y_{1,1} \,,\
{e^{-ih_2} Y_{-}^{(2)}}  = -1/Y_{1,-1} \,,
\\\nonumber
&&{e^{-ih_1} Y_{+}^{(1)}}  = -Y_{2,2} \,,\  {e^{-ih_2}
Y_{+}^{(2)}} =- Y_{2,-2} \,,
\\\nonumber
 &&Y_{Q|vw}^{(1)} = 1/Y_{Q+1,1} \,,\ Y_{Q|vw}^{(2)} = 1/Y_{Q+1,-1}\,,
 \
 Y_{Q|w}^{(1)} = Y_{1,Q+1} \,,\ Y_{Q|w}^{(2)} = Y_{1,-Q-1}\,,~~~~~ \eea
one finds that all the Y-equations discussed above do match the
corresponding ones in the Y-system of \cite{GKV1}.


\section{Conclusions}
In this paper we have derived an infinite set of the TBA equations
for the $\AdS$ mirror theory. There are several obvious questions
to be answered. First, one needs to rigorously establish the
vanishing of Witten's index ($\gamma=\pi$) in the mirror theory,
the latter equals the ground state energy in the original string
model, and see if/how it implies the quantization of the light-cone
momentum of string theory or, equivalently, the quantization of
the temperature  of the mirror model.
 Second, one has to elaborate on the properties of the
dressing phase in the mirror theory. Third, one should find an
analytic continuation of the TBA equations to account for the
excited states. All these questions, of course, are not
independent.

\smallskip

The Y-system we obtained from these equations (under certain
unproved assumptions about the dressing phase!) coincides with
that by \cite{GKV1} only for $u$ taking values  in the interval
$-2<u<2$.  For other values of $u$ one has to assume the validity
of (\ref{KSKi}), and impose the additional condition
(\ref{extrac})  on the Y-functions. The condition, however,  does
not seem to be compatible with the vanishing of the ground state
energy, and if so the Y-system could be valid only for $|u|<2$.

\smallskip

Our brief comparison to the recent results
\cite{Bombardelli:2009ns,GKV2} reveals the following. The TBA
equations we derived in section 3 seem to agree with those of
\cite{Bombardelli:2009ns}. The detailed comparison is difficult to
carry out, however, due to notation differences between the
various sections in \cite{Bombardelli:2009ns}. The simplified form
of the TBA equations we obtained in section 4 does not appear in
\cite{Bombardelli:2009ns,GKV2}. Once again, we see that the
properties of our TBA kernels are such that they admit a localized
Y-system in the interval $u\in (-2,2)$ only. We could not find any
indication of this fact in \cite{Bombardelli:2009ns,GKV2}. Such an
unusual feature arises due to the TBA equation for $Q$-particles
and it is, of course, absent in the Y-system for the homogeneous
Hubbard model. Then, it is assumed in
\cite{Bombardelli:2009ns,GKV2} that the discrete Laplace operator
annihilates the dressing phase kernel. There seems to be no proof
of this assumption in \cite{GKV2}, and in the appendix 2 of
\cite{Bombardelli:2009ns} a proof is given based on the AFS form
\cite{AFS} of the dressing phase. This form, however, is not valid
in the mirror region because the corresponding series is not
convergent.

\smallskip

We also see a certain difference with the findings of \cite{GKV2}.
First of all, none of the TBA kernels in \cite{GKV2} seem to
contain the (square root) of the ratio $x(u+i/gQ)/x(u-i/g Q)$,
while our kernels $K_{\pm}^{Qy}$ do. Further, as was argued in
\cite{AFtba}, the kinematical region of the mirror theory
corresponds to ${\rm Im}\, x^{\pm}<0$ (or ${\rm Im}\, x^{\pm}>0$)
which is apparently different from the choice $|x^+|>1$ and
$|x^-|<1$ indicated in \cite{GKV1,GKV2}. Indeed, the region
$|x^+|>1$ and $|x^-|<1$ is a ``hour-glass'' in Fig.1 of
\cite{AFtba}, while the kinematical region of the mirror theory is
a ``leaf'', where ${\rm Im}\, x^{\pm}<0$ (or ${\rm Im}\,
x^{\pm}>0$). Although both regions do include the physical
momentum of the mirror theory, it is only the leaf which contains
all the bound state solutions of the mirror theory corresponding
to the first and the second BPS families \cite{AFtba}.

\smallskip

In any case, in spite of all the differences and yet to be
justified assumptions, the present effort of deriving the TBA
equations for the $\AdS$ mirror brings us closer towards the final
solution of the AdS/CFT spectral problem.

\section*{Acknowledgements}
The work of G.~A. was supported in part by the RFBR grant
08-01-00281-a, by the grant NSh-672.2006.1, by NWO grant 047017015
and by the INTAS contract 03-51-6346. The work of S.F. was
supported in part by the Science Foundation Ireland under Grant
No. 07/RFP/PHYF104.

\section{Appendices}
\subsection{Mirror dispersion and parametrizations}\la{app:disp}
In this paper we express all the quantities of interest in terms
of the following function of the $u$-plane rapidity variable
\bea\la{xu} x(u)&=& {1\ov 2}\left( u - i \sqrt{4 -
u^2}\right)\,,\quad {\rm Im}(x(u))<0 \ \ {\rm for\  any}\   u\in
{\mathbb C}\,, \eea with the cuts in the $u$-plane running from
$\pm\infty$ to $\pm2$ along the real  lines. Our choice of the
square root cut agrees with the one used in Mathematica: it goes
along the negative semi-axes.
 One can check that with this choice of the cuts the imaginary part of  $x(u)$ is negative  for any  $u\in {\mathbb C}$.
  According to \cite{AFtba}, this function maps the $u$-plane with the cuts onto the physical region of the mirror theory.
  To describe bound states of the mirror model, one should also add either the both lower or both upper edges of the cuts to the $u$-plane \cite{AFtba}. This breaks the parity invariance of the model.

The function $x(u)$ obviously satisfies the condition
$$x(u)+{1\ov x(u)}=u,$$
and  also the following relations
\bea\nonumber
 \ x(-u)=-\frac{1}{x(u)}\, , \quad
(x(u))^*=\frac{1}{x(u^*)}\,.
\eea

The variables $x^{Q\pm}(u)$ used in \cite{AFsh} are expressed through $x(u)$ as follows
\bea\la{xqpm}
x^{Q+}(u) = x(u+ {iQ\ov g})\,,\quad x^{Q-}(u) = x(u- {iQ\ov g})\,,
\eea
where the parameter $g$ is  the
string tension, and it is related to the 't Hooft coupling $\lambda$
of the dual gauge theory as $g={\sqrt\lambda\ov 2\pi}$.

\smallskip

The momentum $\tp^Q$, and the energy $\tH^Q$ of a mirror $Q$-particle are expressed in
terms of $x(u)$ as follows
\bea\la{pu}
\tp^Q(u)&=& g\, x(u-{iQ\ov g}) -g\, x(u+{iQ\ov g})+ i Q \,,\\
\la{Hu}
\tH^Q(u)&=&\log {x(u-{iQ\ov g})\ov x(u+{iQ\ov g})} = 2\,
{\rm arcsinh}\Big( {1\ov 2g}\sqrt{Q^2+\tp^2}\Big)\,,~~~~
\eea
and the momentum is real, and the energy is positive for real values of $u$.
They satisfy the relations
 \bea\nonumber
\tp(-u)=-\tp(u)\, , \quad
(\tp(u))^*=\tp(u^*)\,,\qquad \tH(-u)=\tH(u)\, , \quad
(\tH(u))^*=\tH(u^*)\,.\eea


\subsection{Kernels}\la{app:kern}
Let us introduce the following kernels
\bea\la{Kuv}
K(u,v)=\frac{ 1}{2 \pi i}\,\frac{ \sqrt{4-v^2}}{\sqrt{4-u^2}}\, {1\ov u-v} \, ,~~~
\eea
and  (c.f. \cite{GKV})
\bea\la{KMu}
K_M(u)=\frac{1}{2\pi
i}\frac{d}{du}\log\Big(\frac{u-i{M\ov g}}{u+i{M\ov
g}}\Big) ={1\ov \pi }\, {gM\ov M^2+g^2u^2}\,,\quad -\infty\le M\le \infty\,.
\eea
The Fourier transform of the kernel is
\bea
\hK_M(\om) = \int_{-\infty}^\infty {\rm d}u\, e^{i \om u} K_M(u) ={\rm sign}(M)\, e^{-|M\om|/g}\,,
\eea
and therefore
\bea
\int_{-\infty}^\infty\, {\rm d}u\, K_M(u-u') = {\rm sign}(M)\quad  {\rm for\ any\ }  u' \,.
\eea
Then the kernels $K^{Qy}_\pm$ are related to them  as follows
\bea\nonumber &&K^{Qy}_-(u,v) +
K^{Qy}_+(u,v)=K_{Q}(u-v)\, ,\\\la{KQyuv}
 && K_{Qy}(u,v)\equiv K^{Qy}_-(u,v) -
K^{Qy}_+(u,v)=K(u-{i\ov g}Q,v)-K(u+{i\ov g}Q,v)\, ,~~~~~~
\eea
and the kernels $K^{yQ}_\pm$   as
\bea &&K^{yQ}_-(u,v) -
K^{yQ}_+(u,v)=K_Q(u-v)\, ,\\
&& K_{yQ}(u,v)\equiv K^{yQ}_-(u,v) + K^{yQ}_+(u,v)= \la{KyQuv}
K(u,v+{i\ov g}Q)-K(u,v-{i\ov g}Q)\, .~~~~~~
\eea
The important property of the kernel $K_{yM'}^-$ and $K_{yM'}^+$ is that they are positive for $-2\le u'\le 2$, and satisfy
\bea
\int_{-2}^2\, {\rm d}u\,K_{yQ}(u,u') = 1\quad  {\rm for\ any\ } u'\,.
\eea
We also have
\bea
K^{yM}_{vw}(u,u')\equiv K_M(u-u')\,.
\eea
Then the kernel $K^{MN}_{vv}(u,u')$ can be expressed in terms of $K_M$ as follows
\bea
&& K^{MN}_{vv}(u,u')\equiv K_{MN}(u-u')\,,\\\nonumber
 && K_{MN}(u) =K_{M+N}(u)+K_{N-M}(u)+2\sum_{j=1}^{M-1}K_{N-M+2j}(u)\,,\quad M,N> 0\,.
\eea
The Fourier transform of the kernel is
\bea
\hK_{MN}(\om) = \coth\left({|\om|\ov g}\right)\left( e^{-|M-N||\om|/g} -  e^{-(M+N)|\om|/g} \right) - \delta_{MN}\,,\quad M,N> 0\,.~~~~~~
\eea
The inverse of the kernel is
\bea\nonumber
&&\left( \hK(\om)  + 1\right)_{MN}^{-1} =  \delta_{MN} - \hat{s}(\om)  \left( \delta_{M+1,N}+ \delta_{M-1,N}\right)\,,\quad \hat{s}(\om) = {1\ov 2\cosh {\om\ov g}}\,,~~~~~~\\
\la{invK}
&&\sum_{N=1}^\infty\left( \hK(\om)  + 1\right)_{MN}^{-1} \left( \hK_{NQ}(\om)  + \delta_{NQ}\right)= \delta_{MQ}\,.~~~~~
\eea
Inverse Fourier of $\hat{s}(\om)$ is
\bea\la{shu}
s(u)= {1\ov 2\pi} \int_{-\infty}^\infty {\rm d}\om\, e^{-i \om u} \hat{s}(\om) ={g\ov 4\cosh {g\pi u\ov 2}}\,.
\eea
There is this interesting identity
\bea\la{kmnkm}
\sum_{N=1}^\infty\left( \hK_{MN}(\om)  + \delta_{MN}\right)^{-1} \,\hK_{N}(\om) = \hat{s}(\om)\,\delta_{M1}\,.
\eea
One can show that
\bea
K_{vwx}^{NN'}(u,u')&=& {1\ov 2}K_{NN'}(u-u') +  {\cal K}_1^{NN'}(u,u')\la{kvwxNN3}
\,, ~~~~~
\eea
where
\bea\nonumber
 {\cal K}_1^{NN'}(u,u') = &+&
{1\ov 2}\left(K(u+{i\ov g}N,u'+{i\ov g}N') -K(u-{i\ov g}N,u'-{i\ov g}N')\right)~~~~~~~ \\\la{kvwxNN4}
&+&
{1\ov 2}\left(K(u-{i\ov g}N,u'+{i\ov g}N') -K(u+{i\ov g}N,u'-{i\ov g}N') \right)\,,
\eea
and
\bea\la{kxvNN3}
K_{xv}^{NN'}(u,u')={1\ov 2}K_{NN'}(u-u')+{\cal K}_2^{NN'}(u,u')\,. ~~~~~
\eea
where
\bea\nonumber
{\cal K}_2^{NN'}(u,u')&=& -
{1\ov 2}\left(K(u+{i\ov g}N,u'+{i\ov g}N') -K(u-{i\ov g}N,u'-{i\ov g}N')\right)~~~~~~~ \\\la{kxvNN4}
&+&
{1\ov 2}\left(K(u-{i\ov g}N,u'+{i\ov g}N') -K(u+{i\ov g}N,u'-{i\ov g}N') \right)\,. ~~~~~
\eea

\medskip

Below we list various identities necessary to simplify the TBA equations. The summation over repeated indices from 1 to $\infty$ is assumed
\bea\la{kmnkm2}
\left( K  + 1\right)_{MN}^{-1} \star K_{N} = s\,\delta_{M1}\,.
\eea
\bea\la{KyQKia}
 K_{yQ'}\star (K + 1)^{-1}_{Q'Q} &=&\delta_{Q1}(K_{y1}- K_{y2}\star s) = \delta_{Q1}(s + 2 \check{K}\star s)\,,~~~~~~~~~\\\nonumber
(\check{K}\star s)(u,v) &=&   \int_{-\infty}^{\infty}{\rm d}u'\, \check{K}(u,u')s(u'-v)=\Big( \int_{-\infty}^{-2} -  \int^{\infty}_{2} \Big){\rm d}u'\, \bar{K}(u,u')s(u'-v)\\
\nonumber
&=&
 \pm  \Big( \int_{-\infty}^{-2} +  \int^{\infty}_{2} \Big){\rm d}u'\, K(u,u'\pm i\e)s(u'-v)\,,
\eea
where we introduce the kernel
\bea\la{cKuva}
\check{K}(u,v)= \bar{K}(u,v)\left(\theta(-v-2) -\theta(v-2)\right)\,,\quad
\bar{K}(u,v)=
\frac{ 1}{2 \pi }\,\frac{ \sqrt{v^2-4}}{\sqrt{4-u^2}}\, {1\ov u-v}   \,,  ~~~~~~~
\eea
and $\theta( u)$ is the standard unit step function.
\bea
&&(K + 1)^{-1}_{QQ'}\star K_{Q'y} =\delta_{Q1}(K_{1y}-s\star K_{2y})\,,\\
&&(K_{1y}-s\star K_{2y})(u,v) = s(u,v) \pm 2 \Big( \int_{-\infty}^{-2} +  \int^{\infty}_{2} \Big){\rm d}u'\, s(u-u') K(u'\mp i\e,v)\,.  ~~~~~~~
\eea
\bea\la{Kxvqq}
K^{QQ''}_{xv}\star \left( K +1\right)_{Q''Q'} ^{-1} &=&
 \delta_{Q-1,Q'}s +\delta_{Q'1}  K_{Qy}\star s \,,~~~~~
\eea \bea &&\hspace{-.1cm}(K + 1)^{-1}_{QQ''}\star K^{Q''Q'}_{xv}
=\delta_{Q-1,Q'}s +\delta_{Q1}(K_{xv}^{1Q'}-s\star
K_{xv}^{2Q'})\,,\\ && K_{xv}^{1Q'}-s\star K_{xv}^{2Q'} = \pm \Big(
\int_{-\infty}^{-2} +  \int^{\infty}_{2} \Big){\rm d}u'\,
\nonumber\\
&&~~~~~~~~~~\times s(u-u') (K(u'\mp i\e,v+{i\ov g}Q')+K(u'\mp i\e,v-{i\ov
g}Q'))\,. ~~~~~~~ \eea \bea\la{eqq2a}
 \left( K  + 1\right)_{MN'}^{-1}\star K^{N'Q}_{vwx}&=&
 \delta_{M+1,Q} s +\delta_{M1} s\star K_{yQ} \,.~~~~~
\eea
\bea\nonumber
&&\hspace{-.5cm}
K^{MQ'}_{vwx}\star (K + 1)^{-1}_{Q'Q}  =\delta_{M+1,Q}s +\delta_{1Q}(K_{vwx}^{M1}-K_{vwx}^{M2}\star s)=\delta_{M+1,Q}s +\delta_{1Q}\check{K}_{M}\star s\,,\\\la{KvwKia}
&&\hspace{-.5cm}
(\check{K}_{M}\star s)(u,v)= \int_{-\infty}^{\infty}{\rm d}u'\, \check{K}_M(u,u')s(u'-v)\\\nonumber
&&~~~~~~~~~=  \Big( \int_{-\infty}^{-2} -  \int^{\infty}_{2} \Big){\rm d}u'\, (\bar{K}(u+{i\ov g}M,u') + \bar{K}(u-{i\ov g}M,u'))s(u'-v)\\ \nonumber
&&~~~~~~~~~ =\pm  \Big( \int_{-\infty}^{-2} +  \int^{\infty}_{2} \Big){\rm d}u'\, (K(u+{i\ov g}M,u'\pm i\e)+K(u-{i\ov g}M,u'\pm i\e)) s(u'-v)\,,
\eea
where
\bea\la{cKMuva}
\check{K}_M (u,v)= (\bar{K}(u+{i\ov g}M,v) + \bar{K}(u-{i\ov g}M,v))\left(\theta(-v-2) -\theta(v-2)\right)\,.
\eea
The kernels $\check{K}(u,v)$ and $\check{K}_M(u,v)$ obviously vanish for $|v|<2$.
\bea\la{EQKia}
 \tH_{Q'}\star (K + 1)^{-1}_{Q'Q} &=&\delta_{Q1}( \tH_{1} - \tH_{2}\star s) = \delta_{Q1} \,\check{\cal E}\star s\,,~~~~~~~~~\\\nonumber
(\check{\cal E}\star s)(v) &=&   \int_{-\infty}^{\infty}{\rm d}u'\, \check{\cal E}(u')s(u'-v)\\
\nonumber
&=&
2( \int_{-\infty}^{-2} -  \int^{\infty}_{2} \Big){\rm d}u'\, \log \left(\frac{1}{2}
   \left|u'-\sqrt{u'^2
   -4}\, \right|\right)s(u'-v)\,,
\eea
where we introduce the function
\bea\la{cEua}
\check{\cal E}(u)= 2\log \left(\frac{1}{2}
   \left|u-\sqrt{u^2
   -4}\, \right|\right)\left(\theta(-u-2) -\theta(u-2)\right)\,,
\eea
\bea
&&(K + 1)^{-1}_{QQ'}\star {d\tp^{Q'}\ov du} =\delta_{Q1}( {d\tp^{1}\ov du}-s \star {d\tp^{2}\ov du})\,,\\
&&({d\tp^{1}\ov du}-s\star {d\tp^{2}\ov du})(u) = \pm  \Big( \int_{-\infty}^{-2} +  \int^{\infty}_{2} \Big){\rm d}u'\, s(u-u') {ig u' \ov \sqrt{4-(u'\mp i\e)^2}}\\
&&~~~~~~~~~~~~~~~~~~~~~~~ =\Big( \int^{\infty}_{2} -  \int_{-\infty}^{-2}  \Big){\rm d}u'\, s(u-u') {g u' \ov \sqrt{u'^2-4}}\,,  ~~~~~~~
\eea
\bea
 K_{Q}\star K_{yQ'}&=&\int_{-2}^2 {\rm d}u'\, K_{Q}(u-u')
 K_{yQ'}(u',v)=\\
 \nonumber
 &=&{\cal
 K}_1^{QQ'}(u,v)-\frac{1}{2}K_{Q'-Q}(u-v)+\frac{1}{2}K_{Q'+Q}(u-v)\, .
\eea
 \bea
 K_{Qy}\star K_{Q'}&=&\int_{-2}^2 {\rm d}u'\, K_{Qy}(u,u')
 K_{Q'}(u'-v)=  \\
 \nonumber
&=&{\cal K}_2^{QQ'}(u,v)+\frac{1}{2}K_{Q'-Q}(u-v)+\frac{1}{2}K_{Q'+Q}(u-v)\, . \eea
The following formula holds for $|v| <2$
\bea
2{\cal K}_2^{Q1}\star s(u,v) = K_{Qy}(u,v) \pm 2 \Big( \int_{-\infty}^{-2} +  \int^{\infty}_{2} \Big){\rm d}u'\, K_{Qy}(u,u'\mp i\e) s(u'-v+{i\ov g})  \,.~~~~
\eea
Finally
\bea
\left(  K_{Qy} + K_Q\right)\star s^{-1}=2 K_{xv}^{Q1} + 2\delta_{Q1}\,,\quad K_M\star s^{-1} = K_{M1} + \delta_{M1}\,.
\eea


\subsection{Simplified TBA equations and Y-system}
Here for reader's convenience we list all the simplified TBA  equations and also the Y-system equations. Recall that we introduce the Y-functions related to
the pseudo-energies as
\bea\la{Yfa}
Y_{Q} = e^{-\e_{Q}}\,,\quad Y_{M|vw}^{(\a)} = e^{\e_{M|vw}^{(\a)}}\,,\quad Y_{M|w}^{(\a)} = e^{\e_{M|w}^{(\a)}}\,,\quad Y_\pm^{(\a)} =e^{\e_{y^\pm}^{(\a)}}\,,\quad \a=1,2\,,
\eea
and
assume summation over repeated indices.

The simplified TBA equations for the Y-functions
take the following form
\begin{itemize}
\item $M|w$-strings: $\ M\ge 1\ $, $Y_{0|w}^{(\a)}=0$
\bea\la{Yforwa}
\log Y_{M|w}^{(\a)}=  \log(1 +  Y_{M-1|w}^{(\a)} )(1 +  Y_{M+1|w}^{(\a)} )\star s
+\delta_{M1}\, \log{1-{e^{ih_\a}\ov Y_-^{(\a)}}\ov 1-{e^{ih_\a}\ov Y_+^{(\a)}} }\star s\,~~~~~
\eea

\item $M|vw$-strings: $\ M\ge 1\ $, $Y_{0|vw}^{(\a)}=0$
\bea\la{Yforvw2a}
\hspace{-0.3cm}\log Y_{M|vw}^{(\a)}&=&\log(1 +  Y_{M-1|vw}^{(\a)} )(1 +  Y_{M+1|vw}^{(\a)} )\star s\\\nonumber
&-& \log(1 +  Y_{M+1})\star s+\delta_{M1}  \log{1-e^{-ih_\a}Y_-^{(\a)}\ov 1-e^{-ih_\a}Y_+^{(\a)}}\star s\,~~~~~
\eea

\item $y$-particles
\bea\la{Yfory1a}
&&\log {Y_+^{(\a)} \ov Y_-^{(\a)}}=  \log(1 +  Y_{Q})\star K_{Qy} \,,\\
\la{Yfory2a}
&&\log {Y_+^{(\a)}  Y_-^{(\a)}} =  - \log\left(1+Y_Q \right)\star K_Q+2\log {1+{1\ov Y_{M|vw}^{(\a)}}
 \ov1+{1\ov Y_{M|w}^{(\a)}}}\star K_M\,
\eea

\item $Q$-particles for $Q\ge 2$
\bea\nonumber
-\log Y_{Q}+ \log(Y_{Q-1}Y_{Q+1})\star s&=&\log{(1 +  Y_{Q-1} )(1 +  Y_{Q+1} ) \ov \left(1 +  {1\ov Y_{Q-1|vw}^{(1)}} \right)\left(1 +  {1\ov Y_{Q-1|vw}^{(2)}} \right)}\star  s\\\la{YforQ2a}
&+&2\log\left(1+Y_{Q'} \right)\star K^\Sigma_{Q'Q''}\star (K + 1)^{-1}_{Q''Q} \,~~~~~~~
\eea

\item $Q=1$-particle
\bea
\la{YforQ1a}
-\log Y_{1}&=& \log(1 +  {1\ov Y_{2}} )\star s
-\log\left(1-{e^{ih_1} \ov Y_{-}^{(1)}} \right)\left(1-{e^{ih_2} \ov Y_{-}^{(2)}}\right)\star s \\\nonumber
&-&  \log\left(1-{e^{ih_1} \ov Y_{-}^{(1)}} \right)\left(1-{e^{ih_2} \ov Y_{-}^{(2)}}\right)
\left(1-{e^{ih_1} \ov Y_{+}^{(1)}} \right)\left(1-{e^{ih_2} \ov Y_{+}^{(2)}}\right)
\star \check{K}\star s
\\
\nonumber&-& \log\left(1 +  {1\ov Y_{M|vw}^{(1)}} \right)\left(1 +  {1\ov Y_{M|vw}^{(2)}} \right)\star \check{K}_M\star s
\\\nonumber
&+&2\log\left(1+Y_{Q} \right)\star K^\Sigma_{QQ'}\star (K + 1)^{-1}_{Q'1} +L\, \check{\cal E}\star s   \,~~~~~
\eea

\end{itemize}
The energy of the ground state of the light-cone gauge-fixed string theory on $\AdS$ is expressed through the Y-functions as follows
 \bea
\la{energyLa} E_\g(L) &=&-\int {\rm d}u\, \sum_{Q=1}^\infty{1\ov
2\pi}{d\tp^Q\ov du}\log\left(1+Y_Q\right)\,. \eea

\medskip

With the notation $Y^{\pm}(u)\equiv Y(u\pm {i\ov g}\mp i0)$ for any Y-function, the Y-system equations for $u\in [-2,2]$
take  the following form
\begin{itemize}
\item $M|w$-strings
\bea\la{YwMa}
Y_{M|w}^{(\a)+}\,Y_{M|w}^{(\a)-} &=& \left( 1+Y_{M-1|w}^{(\a)} \right)\left( 1+Y_{M+1|w}^{(\a)} \right)  \quad {\rm if}\ \ M\ge 2\,,\\\la{Yw1a}
Y_{1|w}^{(\a)+}\,Y_{1|w}^{(\a)-} &=& \left( 1+Y_{2|w}^{(\a)} \right){1-{e^{ih_\a}\ov Y_-^{(\a)}}\ov 1-{e^{ih_\a}\ov Y_+^{(\a)}} } \,
\eea

\item $M|vw$-strings
\bea\la{YvwMa}
Y_{M|vw}^{(\a)+}\, Y_{M|vw}^{(\a)-} &=& \left( 1+Y_{M-1|vw}^{(\a)} \right)\left( 1+Y_{M+1|vw}^{(\a)} \right)\, {1\ov 1+Y_{M+1}}  \quad {\rm if}\ \ M\ge 2\,,~~~~~~~\\\la{Yvw1a}
Y_{1|vw}^{(\a)+}\, Y_{1|vw}^{(\a)-} &=& { 1+Y_{2|vw}^{(\a)} \ov 1+Y_{2}} \, {1-e^{-ih_\a}Y_-^{(\a)}\ov 1-e^{-ih_\a}Y_+^{(\a)}} \,
\eea

\item $y^-$-particles
\bea\la{tbaym4a}
Y_{-}^{(\a)+}\,Y_{-}^{(\a)-}  &=& {1+ Y_{1|vw}^{(\a)} \ov1+ Y_{1|w}^{(\a)}} \, {1\ov 1+Y_1 }
\,
\eea
\item $y^+$-particles
\bea
\la{tbayp2a}
\log Y_{+}^{(\a)+}\,Y_{+}^{(\a)-}
&=&2 \log\left(1+Y_Q \right)\star {\cal K}_{2}^{Q1}  +\log {1+ Y_{1|vw}^{(\a)} \ov1+ Y_{1|w}^{(\a)}}
 \,
\eea
\item $Q$-particles for $Q\ge 2$
\bea\la{YQQa}
{Y_{Q}^{+}\, Y_{Q}^{-}\ov Y_{Q-1}Y_{Q+1}} &=& {\left(1 +  {1\ov Y_{Q-1|vw}^{(1)}} \right)
\left(1 +  {1\ov Y_{Q-1|vw}^{(2)}} \right)\ov \left( 1+Y_{Q-1} \right)\left( 1+Y_{Q+1} \right) } \,
\eea
\item $Q=1$-particle
\bea\la{YQ1a}
{Y_{1}^{+}\, Y_{1}^{-}\ov Y_{2}} &=& {\left(1 -  {e^{ih_1}\ov Y_{-}^{(1)}} \right)
\left(1 -  {e^{ih_2}\ov Y_{-}^{(2)}} \right)\ov  1+Y_2 }\, ,\eea
where we assume the validity of the formula
\bea\la{KSKia}
K^\Sigma_{Q'Q''}\star (K + 1)^{-1}_{Q''Q} \stackrel {?}{=}\delta_{1Q}\check{K}_{Q'}^\Sigma\star s\,.
\eea

\end{itemize}


\end{document}